\documentclass[10pt,a4paper,twoside]{article}

\usepackage{indentfirst}			
\usepackage{graphicx}				
\usepackage{amsgen,amsfonts,amssymb,amsbsy,amsmath}	

\newenvironment{resum}{\begin{quote}\small}{\end{quote}}

\newcommand{\bfsf}[1]{\textsf{\textbf{#1}}}

\begin{document}

\thispagestyle{plain}		

\begin{center}


{\LARGE\bfsf{New families of singularity-free cosmological models}}

\bigskip


\textbf{L. Fern\'andez-Jambrina}$^1$ and \textbf{L.M. Gonz\'alez-Romero}$^2$


$^1$\textsl{Universidad Polit\'ecnica de Madrid, Spain.} \\
$^2$\textsl{Universidad Complutense de Madrid, Spain.}

\end{center}

\medskip


\begin{resum}
In this talk we extend a family of geodesically complete $G_{2}$ stiff fluid 
cosmological models to the case in which the velocity of the fluid is 
not orthogonal to the gradient of the transitivity surface element.
\end{resum}

\bigskip


\section{Introduction}

During the previous decade the interest for singularity-free 
cosmological models was supported by the publication of the first 
known geodesically complete perfect fluid cosmology \cite{seno}. 
Until then such possibility had been overlooked due to the restrictive 
conditions imposed by singularity theorems \cite{HE}, 
\cite{beem}. These theorems required physically reasonable 
restrictions such as energy, causality and generic conditions, but 
they also imposed the existence of certain trapped sets, such as compact achronal
sets without edge or closed trapped surfaces, which were not so  
obvious. Namely this latter condition is the one which has 
been used to avoid the formation of singularities. More details about 
geodesic completeness of the Senovilla spacetime may be found in 
\cite{chinea}.

Since 1990, the number of new cosmological models which were singularity-free did 
not increase very much, most of them within the 
framework of $G_{2}$ orthogonally transitive spacetimes \cite{grg}. But in 2002 
a family of stiff fluid cosmologies depending on two almost arbitrary 
functions was shown to be geodesically complete \cite{wide}. It is so far the 
largest family of singularity-free perfect fluid cosmologies. It was 
obtained requiring that the velocity of the fluid should be orthogonal 
to the gradient of the transitivity surface element at every point of 
the spacetime. In this talk we 
show that this restriction may be removed and thereby the family of 
singularity-free cosmologies is enlarged.

In the next section we show the Einstein equations for $G_{2}$ 
orthogonally transitive stiff fluid spacetimes and simplify them so 
that the analysis of the geodesics may be carried out conveniently. In Section 3 
regularity theorems \cite{manolo} are used to derive conditions on the 
spacetimes to be geodesically complete.

\section{Einstein equations for $G_{2}$ stiff fluid cosmologies\label{equations}}

We shall write the Einstein equations for the metric of a spacetime 
endowed with an Abelian orthogonally transitive group of 
isometries $G_{2}$ acting on spacelike surfaces. Except for some 
spherically symmetric spacetimes, every other singularity-free 
cosmology belongs to that group. We shall also require that the metric 
be diagonal.

We choose a coordinate chart in which the metric is isotropic in the 
non-ignorable coordinates, $t$, $r$. The coordinates $z$ and $\phi$ 
are adapted to the generators of the isometry group and therefore 
metric functions do not depend on them,

\begin{equation}
g=e^{2K}(-dt^2+dr^2)+e^{-2U}dz^2+\rho^2e^{2U}d\phi^2.\label{metric}
\end{equation}

The range of the coordinates is the standard,
\begin{equation}
    -\infty<t,z<\infty,\ 0<r<\infty,\ 0<\phi<2\pi,
\end{equation}
for cylindric symmetry.

The matter content of the spacetime is a perfect fluid,

\begin{equation}
    T=\mu u\otimes u+p\,(g+u\otimes u),
\end{equation} 
for which pressure, $p$, and  energy density, $\mu$, are equal. 
The velocity of the fluid $u$ is parametrized with the help of a 
function $\xi$,

\begin{equation}
    u=-e^{K}\,(\cosh \xi dt +\sinh\xi dr),
\end{equation} 
which allows to write the Einstein equations for a stiff perfect fluid as

\begin{subequations}
\begin{eqnarray}
    && U_{tt}-U_{rr}+\frac{1}{\rho}(U_{t}\rho_{t}-U_{r}\rho_{r})=0,\label{U1}
    \\
    && \rho_{tt}-\rho_{rr}=0,\label{rho1}
    \\
    && 
    \frac{K_{t}\rho_{r}+K_{r}\rho_{t}}{\rho}=\frac{\rho_{tr}+U_{t}\rho_{r}+U_{r}\rho_{t}}{\rho}+2 
    U_{t}U_{r}+e^{2K} p\sinh2\xi,\label{Kt1}
    \\
    && 
    \frac{K_{t}\rho_{t}+K_{r}\rho_{r}}{\rho}=\frac{\rho_{tt}+\rho_{rr}}{2\rho}+
    \frac{U_{t}\rho_{t}+U_{r}\rho_{r}}{\rho}
    +U_{t}^2+U_{r}^2\nonumber\\&&+ e^{2K}p\cosh2\xi,\label{Kr1}
    \\
    && 
    K_{rr}-K_{tt}+\frac{U_{r}\rho_{r}-U_{t}\rho_{t}}{\rho}+U_{r}^2-U_{t}^2=
    p\,e^{2K}, \label{nu1}\\
    && 
    K_{r}-\xi_{t}+\frac{p_{r}}{2p}+\frac{\rho_{t}\cosh\xi-\rho_{r}\sinh\xi}{\rho}
    \sinh\xi=0,\label{p1}
    \\
    && K_{t}-\xi_{r}+\frac{p_{t}}{2p}
    +\frac{\rho_{t}\cosh\xi-\rho_{r}\sinh\xi}{\rho}
    \cosh\xi=0.\label{mu1}
\end{eqnarray}
\end{subequations}

This differential system may be simplified by taking $\rho$ as our 
coordinate $r$. We thereby restrict the solutions to the set for which 
the gradient of the transitivity surface element, $\rho$, is 
spacelike. This is the set where singularity-free spacetimes have been 
found so far.

\begin{subequations}
\begin{eqnarray}
&& U_{tt}-U_{rr}-\frac{U_{r}}{r}=0,\label{U2}
\\ && 
K_{t}=U_{t}+2r U_{t}U_{r}+e^{2K}pr\sinh2\xi,\label{Kt2}
\\ && 
K_{r}=U_{r}+r(U_{t}^2+U_{r}^2)+e^{2K}pr\cosh2\xi,\label{Kr2}
\\ && 
K_{rr}-K_{tt}+\frac{U_{r}}{r}+U_{r}^2-U_{t}^2=pe^{2K}, \label{nu2}
\\
&& 
K_{r}-\xi_{t}+\frac{p_{r}}{2p}-\frac{\sinh^2\xi}{r}=0,\label{p2}
\\
&& K_{t}-\xi_{r}+\frac{p_{t}}{2p}-\frac{\sinh\xi\cosh\xi}{r}=0.
\label{mu2}
\end{eqnarray}
\end{subequations}

We may further write an exact differential,

\begin{equation}\label{exact}
    dH= e^{2K}rp(\sinh 2\xi\,dt+\cosh 2\xi\,dr),
\end{equation}
which provides $\xi$ and $p$ once $K$ is known,

\begin{equation}\label{pdef}
    \tanh 2\xi= \frac{H_{t}}{H_{r}},\qquad 
    |p|=\frac{e^{-2K}}{r}\sqrt{H_{r}^2-H_{t}^2}.
\end{equation}

This allows another simplification of the system of differential equations,

\begin{subequations}
\begin{eqnarray}
    && U_{tt}-U_{rr}-\frac{U_{r}}{r}=0,\label{U3}
    \\ && 
    H_{rr}-H_{tt}=\frac{\sqrt{H_{r}^2-H_{t}^2}}{r}, \label{H}
    \\ && 
    K_{t}=U_{t}+2r U_{t}U_{r}+H_{t},\label{Kt3}
    \\ && 
    K_{r}=U_{r}+r(U_{t}^2+U_{r}^2)+H_{r},\label{Kr3}    
\end{eqnarray}
\end{subequations}
which is formed by an inhomogeneous 2-D wave equation for $U$ and a 
non-linear wave equation for $H$. The metric function $K$ is obtained 
after integrating  a quadrature. It may be checked that this system is 
consistent.

It can be shown that the axis can be made regular just rescaling the 
angular coordinate 
$\phi$. 

Although it is not necessary for our purposes, we point out that
the Wainright-Ince-Marshman formalism \cite{Wain} may be used for 
generating solutions to the non-linear wave equation.

\section{Singularity-free stiff fluid models\label{theorems}}

The simplified system of Einstein equations that we have shown  is 
suitable for the analysis of geodesic completeness. The case of diagonal 
Abelian orthogonally transitive spacetimes  has been thoroughly studied
in \cite{manolo} and we shall make use of those results, which may be 
written in the form of a theorem,

\begin{description}
\item[Theorem:] A diagonal Abelian orthogonally transitive spacetime 
with spacelike orbits endowed with a metric in the form (\ref{metric}) with $C^2$
metric functions  $K,U,\rho$, where $\rho$ has a spacelike gradient, is future causally geodesically complete 
provided that along causal geodesics:
\begin{enumerate}
\item For large values of $t$ and increasing $r$, 
\begin{enumerate}
    \item $(K-U-\ln\rho)_{r}+(K-U-\ln\rho)_{t}\ge 0$, and either 
    $(K-U-\ln\rho)_{r}\ge 0$  or $|(K-U-\ln\rho)_{r}|\lesssim 
    (K-U-\ln\rho)_{r}+(K-U-\ln\rho)_{t}$.
    \item $K_{r}+K_{t}\ge 0$, and either $K_{r}\ge 0$ or 
    $|K_{r}|\lesssim K_{r}+K_{t}$.
    \item $(K+U)_{r}+(K+U)_{t}\ge 0$, and either $(K+U)_{r}\ge 0$ 
    or $|(K+U)_{r}|\lesssim (K+U)_{r}+(K+U)_{t}$.
\end{enumerate}

\item \label{tt} For large values of  $t$, a constant $b$ exists such that

\begin{eqnarray*}\left.\begin{array}{c}K(t,r)-U(t,r)\\2\,K(t,r)\\K(t,r)+U(t,r)+\ln\rho(t,r)
\end{array}\right\}\ge-\ln|t|+b.\end{eqnarray*}

\end{enumerate}\end{description}

For past-pointing geodesics the corresponding theorem is quite 
similar, just exchanging the sign of the time derivatives and imposing 
conditions for small values of $t$, instead of large values. A 
generalization to nondiagonal metrics may be found in \cite{nondiag}.

The results that are drawn from applying these theorems to the stiff 
fluid case may be summarized as follows:

\begin{description}
\item[Theorem 1:] A cylindrical spacetime 
with a stiff perfect fluid as matter content, 
endowed with a metric in the form (\ref{metric}) with $C^2$
metric functions  $K,U,\rho$ is future geodesically complete 
if the gradient of the surface element is spacelike and
\begin{enumerate}
\item For large values of $t$, $U(t,0)\ge -\displaystyle\frac{1}{2}\ln |t|+b$.

\item Either $r^{1-\varepsilon}|U_{r}+ U_{t}|$ or 
$r^{1-\varepsilon}(H_{r}+ H_{t})$ does not tend to zero for large 
values of $t$ and $r$.

\end{enumerate}\end{description}

\begin{description}
\item[Theorem 2:] A cylindrical spacetime 
with a stiff perfect fluid as matter content, 
endowed with a metric in the form (\ref{metric}) with $C^2$
metric functions  $K,U,\rho$ is past geodesically complete 
if the gradient of the surface element is spacelike and
\begin{enumerate}
\item For small values of $t$, $U(t,0)\ge -\displaystyle\frac{1}{2}\ln |t|+b$.

\item Either $r^{1-\varepsilon}|U_{r}- U_{t}|$ or 
$r^{1-\varepsilon}(H_{r}- H_{t})$ does not tend to zero for small
values of $t$ and large values of $r$.

\end{enumerate}\end{description}

It is not much difficult to find metric functions that comply with these 
requirements:

\begin{description}
    \item[Corolary:] A metric with arbitrary $H$ and 
a function $U$ which grows for large $|t|$ and for large $r$ makes the spacetime 
geodesically complete. 
\end{description}

As it is shown in \cite{wide}, it is not difficult to provide 
solutions of the inhomogeneous wave equation in a form amenable to 
check the premises of the theorems. We may consider the initial value 
problem,

\begin{eqnarray}
    && U_{tt}-U_{rr}-\frac{U_{r}}{r}=0,\nonumber\\
    && U(0,r)=f(r),\quad U_{t}(0,r)=g(r),
    \end{eqnarray}
    which can be solved in closed integral form \cite{john},

\begin{eqnarray}
    U&=&U_{f}+U_{g},\nonumber \\
U_{f}(x,y,t)&=& \frac{1}{2\pi}\int_{0}^{2\pi}d\phi\int_0^tdR 
R\frac{g(x+R\cos\phi,y+R\sin\phi)}{\sqrt{t^2-R^2}}\nonumber\\ U_{g}(x,y,t)&=&
\frac{1}{2\pi}\frac{\partial}{\partial 
t}\int_{0}^{2\pi}d\phi\int_0^tdR 
R\frac{f(x+R\cos\phi,y+R\sin\phi)}{\sqrt{t^2-R^2}},
\end{eqnarray} 
for initial data $U(x,y,0)=f(x,y)$, $U_{t}(x,y,0)=g(x,y)$ with circular symmetry.

The term depending respectively on the 
initial data $f$ is even in $t$. On the contrary, $U_{g}$ is odd. 
Since the condition of the corolary is symmetric in $t$, this means 
that either $U_{f}$ and $U_{g}$ must comply the condition separately 
or just $U_{f}$ complies it, but overcoming the term $U_{g}$. 

In 
\cite{wide} it is shown how this can be done for polynomial initial 
data. For instance if $f$ and $g$ are polynomials and the degree of 
$f$ is larger than the one of $g$ and the coefficient of the leading 
term is positive, we generate a geodesically complete spacetime.

\section{Final remarks\label{conclusions}}

In this talk sufficient conditions for an Abelian 
diagonal orthogonally transitive spacetime  with spacelike 
orbits and with a stiff perfect fluid 
as matter content to be geodesically complete have been produced. The 
resulting conditions are simple and can be 
implemented easily on examples. These results generalize previous 
work by the authors. More details will be found elsewhere 
\cite{stiffgen}. 
 
The main consequence of this talk is that geodesically complete 
spacetimes are more abundant than it was thought before. Since stiff 
fluids are a special case of perfect fluids, that may also be 
interpreted as scalar fields, further work is needed and other 
sources and symmetries are to be considered in the future.

\section*{Acknowledgments}
The present work has been supported by Direcci\'on General de
Ense\~nanza Superior Project PB98-0772. The authors wish to thank
 F.J. Chinea,  F. Navarro-L\'erida and  M.J. Pareja 
for valuable discussions.


\end{document}